\documentclass[twoside,english,journal]{IEEEtran}
\usepackage[latin9]{inputenc}
\usepackage{verbatim}

\makeatletter
\usepackage{booktabs}
\usepackage{graphicx}

\title{Configuration Redundancy for Enhanced Reliability in SRAM-based FPGAs}
\author{Raffaele~Giordano, Sabrina~Perrella, Dario~Barbieri, Vincenzo~Izzo, and Alberto~Aloisio
\thanks{"`Accesso Aperto MIUR"'. This work is part of the ROAL project (CINECA grant no. RBSI14JOUV) funded by the Scientific Independence of Young Researchers (SIR) 2014 program of the Italian Ministry of Education, University and Research (MIUR). The institutions which contributed to the results reported in this work are listed below as affiliations of the authors. Corresponding author: R. Giordano (email: rgiordano@na.infn.it)}
\thanks{R. Giordano, S. Perrella D. Barbieri, and A. Aloisio are with INFN and Università degli Studi di Napoli "Federico II", I-80126, Napoli, Italy}
\thanks{V. Izzo is with INFN Sezione di Napoli, I-80126, Napoli, Italy}
}

\makeatother

\usepackage{babel}
\begin{document}
\maketitle 
 \thispagestyle{empty} \pagestyle{empty} \renewcommand{\figurename}{Fig.} \renewcommand{\tablename}{TABLE}

\begin{abstract}
Digital off-detector electronics in trigger and data acquisition systems
of High-Energy Physics experiments is often implemented by means of
SRAM-based FPGAs, which make it possible to achieve reconfigurable,
real-time processing and multi-gigabit serial data transfers. On-detector
usage of such devices is mostly limited by their configuration sensitivity
to radiation-induced upsets, which may alter the programmed routing
paths and configurable elements. 

In this work, we show a new technique for enhancing the usage of SRAM-based
FPGAs also for on-detector applications. We show a demonstrator of
our solution on benchmark designs, including a triple modular redundant
design and a serial link (without redundancy) running at 5 Gbps, implemented
in a Xilinx Kintex-7 FPGA. We performed irradiation tests at Laboratori
Nazionali del Sud (Catania, Italy) with a 62-MeV proton beam. The
results show that our scrubbing technique made it possible to detect
and correct all the radiation-induced upsets after a total fluence
higher than $10^{12}cm^{-2}$. For both the redundant benchmark design
and the serial link, the correct functionality was always restored
after scrubbing the corrupted configuration bits and resetting the
circuit. 
\end{abstract}

\IEEEPARstart{S}{Static} RAM-based Field Programmable Gate Arrays
(SRAM-based FPGAs) \cite{Xilinx_Ultrascale,Altera_Stratix} are widely
used in trigger and data acquisition (TDAQ) systems of High-Energy
Physics (HEP) experiments for implementing fast logic due to their
re-configurability, large real-time processing capabilities and embedded
high-speed serial IOs. However, these devices are sensitive to radiation
effects such as single event upsets (SEUs) \cite{Wirthlin_HiRel,QuinnRadEffectsFPGAs}
in the configuration memory, which may alter the functionality of
the implemented circuit.%
{} There is a strong interest in finding solutions for enhancing the
usage of standard SRAM-based FPGAs also on-detector. Methods based
on triple modular redundancy (TMR) \cite{TMR_SRAM_Based_FPGAs} and
periodic correction of the configuration, i.e. scrubbing \cite{Scrubbing,HybridScrubbing,Effectivness_of_SEU_scrubbing_mitigation_strategies},
are used in order to correct single and multiple bit upsets (SBUs
and MBUs) per memory location (i.e. frame). The reason for coupling
scrubbing to modular redundancy is that fault masking techniques,
such as TMR, require to avoid the accumulation of errors in the FPGA
\cite{TMR_Accumulation2}. In classical scrubbing approaches, the
golden configuration is stored in external radiation-hardened memories
and it is possible to correct any number of MBUs per frame.

Recently, scrubbing techniques based on redundant configuration have
been developed in order to avoid external memories and to correct
MBUs at the same time, a brief discussion of the literature about
this topic follows. 

Although it cannot be classified as a scrubber, the patent disclosed
in \cite{RedundantConfigurationMemory} shows a very interesting FPGA
architecture and design implementation flow aimed at generating redundant
configuration at the bit level. Unfortunately, since it requires a
dedicated FPGA architecture, this approach can be pursued only by
vendors fabricating devices and it cannot be implemented at the user
level. Moreover, it does not allow to detect upsets, but only to mask
them, therefore it is not effective against the accumulation of upsets
and it does allow to log them. 

The patent described in \cite{triple_PLD_patent} and the work presented
in \cite{Self-referenceScrubber} are both based on configuration
redundancy at the device level. Three identical FPGAs implement the
same design and therefore host the same configuration. The main limitations
of this approach are the need for three devices, the increase of the
power consumption by a factor three and the need for an additional
device to perform scrubbing and majority voting of the outputs. 

An interesting, and effective, approach based on configuration redundancy
at the frame level is shown in \cite{Frame_Level_Redundancy} for
a Virtex-5 FPGA. The technique requires a custom design flow, which
is based on the legacy Xilinx ISE tool, its ability to export layouts
in the Xilinx Design Language (XDL) format, and the Rapidsmith \cite{RapidSmith}
academic CAD tool. The Rapidsmith tool is used for replicating the
layout of a module three times, generating three identical subsets
of the configuration, and therefore the Authors exploit modular redundancy
to generate configuration redundancy. In this implementation the scrubbing
logic and a voter for the three modules are implemented in the fabric.
This solution leads to a power consumption increase related to the
additional programmable resources used. However, the impact on power
consumption is milder with respect to solutions based on redundant
devices \cite{LayoutAndRadTolIssues}, where also the device quiescent
power is triple. 

Unfortunately, newer FPGA families, such as the 7-Series, the Ultrascale
or the Ultrascale+, are not supported by the Rapidsmith tool, and
FPGAs of the latest generation are not supported by ISE either. In
addition to that, the new Xilinx CAD tool Vivado, recommended for
designs based on 7-Series onwards, does not support the XDL. New initiatives,
such as \cite{Rapidsmith2}, have been launched for enabling custom
design flows also with the Vivado tool. However, the usage of third-party
layout tools adds up to the complexity of the design flow and it usually
does not allow the designer to choose devices of the latest generation,
since the dedicated support must be implemented for each new FPGA
family. 

A new method for generating redundant configuration at the frame level
has been described in \cite{Redundant_Configuration}. In this case
the used configuration frames content is copied to unused frames after
the initial configuration of the device. There is no dependency on
the layout tools and the solution is exportable to any other device,
provided it includes an interface for reading and writing the internal
configuration. The impact on power consumption is minimal with respect
to the previously cited solutions and in general to TMR-based solutions.
A common aspect of the methods described in \cite{Frame_Level_Redundancy}
and \cite{Redundant_Configuration} is that they both require to find
identical subsets of the FPGA device for hosting the redundant configuration
(or modules). 

This work advances the state of the art in multiple ways. 

We show an enhanced version of the configuration redundancy generation
and scrubbing technique for SRAM-based FPGAs published in \cite{Redundant_Configuration}.
In fact, our new version relaxes the requirement of allocating the
redundant configuration in identical subsets of the FPGA. Moreover,
the presented technique is capable of protecting the configuration
pertaining to basic blocks such as flip-flops, look-up-tables and
routing, but it can also address complex hard macros, such as high-speed
IO transceivers (e.g. the Xilinx GTX). 

We show a demonstrator of our solution on two benchmark designs implemented
in a Xilinx Kintex-7 FPGA: an array of 32-bit counters emulating a
classical mixture of sequential and combinational logic and a serial
link running at 5 Gbps without modular redundancy. For the first benchmark
design we implemented the TMR scheme by means of the Synopsys Synplify
Premier tool \cite{Synplify}.

We present results from irradiation tests performed at Laboratori
Nazionali del Sud (Catania, Italy) with a 62-MeV proton beam with
a total fluence of $3.5\cdot10^{11}cm^{-2}$ for the TMR-protected
design and of $8.7\cdot10^{11}cm^{-2}$ for the serial link. We split
our test into several runs, each corresponding to a fluence in the
range from $2.9\cdot10^{10}cm^{-2}$ to $2.9\cdot10^{11}cm^{-2}$.
We used a software implementation of the scrubber running on a small
form factor personal computer interfaced to the device under test
via JTAG. During irradiation of the FPGA the scrubber was active in
order to remove radiation-induced upsets as soon as they were detected.
For each upset we logged a time stamp, the frame address, the bit
offset and the upset plarity ($0\rightarrow1$ or $1\rightarrow0$).
We have also been measuring the power consumption of the FPGA, separately
at its power domains, and the functionality of the selected benchmark
design.

Our results show that our scrubbing technique made it possible to
detect and correct all the radiation-induced upsets. For both the
TMR-protected firmware and the serial link, the correct functionality
was always restored after scrubbing the corrupted configuration bits
and resetting the circuit. However, the TMR-protected firmware has
shown a significantly lower number of failures with respect to the
serial link. We measured some single event functional interrupts (SEFIs)
related to the readback of block RAMs (BRAMs), where 128 bits at the
same bit offset in logically adjacent frames were flipped. The proton
to upset cross section for the entire device has been measured to
be $\sigma_{dev}=1.0\cdot10^{-7}cm^{2}$, while the proton to upset
cross section per bit has been measured to be $\sigma_{bit}=4.4\cdot10^{-15}cm^{2}$
and the proton to BRAM SEFI cross section has been measured to be
$\sigma_{BRAM}=4.6\cdot10^{-11}cm^{2}$. It is important to note that
$\sigma_{BRAM}$ is more than three order of magnitudes lower than
$\sigma_{dev}$, i.e. SEFIs are very rare events. SEFIs could not
be removed by scrubbing, as they require to power cycle the device,
but they did not impact the operation of our designs.

\end{document}